# Rejoinder: Matched Pairs and the Future of Cluster-Randomized Experiments

Kosuke Imai, Gary King and Clayton Nall

## 1. INTRODUCTION

We are grateful to our four discussants for their agreement with and contributions to the central points in our article (Imai et al. (2009b)). As Zhang and Small (2009) write, "[our article] present[s] convincing evidence that the matched pair design, when accompanied with good inference methods, is more powerful than the unmatched pair design and should be used routinely." And, as they put it, Hill and Scott (2009) "do not take issue with [our article's] provocative assertion that one should pair-match in cluster randomized trials 'whenever feasible.'" Whether denominated in terms of research dollars saved, or additional knowledge learned for the same expenditure, the advantages in any one research project of switching standard experimental protocols from complete randomization to a matched pair designs (along with the accompanying new statistical methods) can be considerable.

In the two sections that follow, we address our discussants' points regarding ways to pair clusters (Section 2) and the costs and benefits of design- and model-based estimation (Section 3). But first we offer a sense of how many experiments across fields of inquiry can be improved in the ways we discuss in our article. We do this by collecting data from the last 106 cluster-randomized experiments published in 27 leading journals in medicine, public health, political science, economics, and education. We then counted how many experiments used complete randomization, blocking (on some but not all pre-treatment information), or pair-matching—which respectively exploit none, some and all of the available pre-randomization covariate information. Table 1 gives a summary. Overall, only 19% of cluster-randomized experiments used pair-matching, which means that 81% left at least some pre-randomization covariate information on the table. Indeed, almost 60% of these experiments used complete randomization and so took no advantage of the information in pre-treatment covariates. The table conveys that there is some variation in these figures across fields, but in no field is the use of pair matching in cluster-randomized designs very high, and it never occurs in even as many as 30% of published experiments. Administrative constraints may have prevented some of these experiments from being pair matched, but as using this information involves no modeling risks, the opportunities for improving experimental research across many fields of inquiry seem quite substantial.

## 2. HOW TO CONSTRUCT MATCHED PAIRS

Zhang and Small (2009) offer some creative ideas on how to construct matched pairs based on minimizing the total (i.e., across pairs) Mahalanobis-based distance metric, which is referred to as an "optimal" method. This procedure can be useful in many situations, and will usually be superior to Mahalanobis-based matching methods that do not consider imbalances for all pairs simultaneously.

This technique, of course, is not always appropriate. For example, the procedure assumes that Mahalanobis distances make sense for the input data, which means that the variance matrix which scales the distances is known or can be estimated, and

*Kosuke Imai is Assistant Professor, Department of Politics, Princeton University, Princeton, New Jersey 08544, USA URL: http://imai.princeton.edu; e-mail: kimai@princeton.edu. Gary King is Albert J. Weatherhead III University Professor, Institute for Quantitative Social Science, Harvard University, 1737 Cambridge St., Cambridge, Massachusetts 02138, USA URL: http://GKing.Harvard.edu; e-mail: King@Harvard.edu. Clayton Nall is Ph.D. Candidate, Department of Government, Institute for Quantitative Social Science, Harvard University, 1737 Cambridge St., Cambridge, Massachusetts 02138, USA e-mail: nall@fas.harvard.edu.*

TABLE 1
*Percent of recent cluster-randomized experiments in each of four research fields using unblocked, blocked (on a subset of available pre-treatment covariates) or pair-matched designs*

| Field | Amount of pre-randomization design information used | | | $N$ |
|---|---|---|---|---|
| | **None** (Unblocked) | **Some** (Blocked) | **All** (Pair-matched) | |
| Medicine and public health | 56.2% | 20.5% | 23.3% | 73 |
| Political science | 71.4 | 23.8 | 4.8 | 21 |
| Economics | 42.9 | 28.6 | 28.6 | 7 |
| Education | 80.0 | 20.0 | 0.0 | 5 |
| Total | 59.4 | 21.7 | 18.9 | 106 |

Row totals may not add to 100% due to rounding. For details on these data, see the Appendix.

that the input variables are close to normal. Perhaps even more importantly, the procedure maps all the distances to a scalar to measure balance; this assumes that the researcher is willing to reduce balance within pairs for some pre-treatment variables in order to achieve a larger improvement for other variables. However, if the set of variables having its balance reduced has a bigger impact on the outcome than the other set, then the trade-off implied by the distance metric would be ill advised. One way to avoid these trade-offs is to use a matching method without a scalar balance metric, such as "coarsened exact matching" which guarantees that the maximum possible imbalance for each variable is set by ex ante user choice (Iacus et al. (2008)).

Our qualifications here are minor, of course, as most versions of pair matching with a good choice of pre-treatment variables would normally represent a tremendous improvement over a complete randomization design with respect to bias, power, efficiency, and robustness. And Zhang and Small's point is clearly correct that one can often do better by considering balance on all pairs simultaneously in the context of scalar distance-based balancing.

Finally, we note that constructing matched pairs in experimental work is similar to the problem of matching in observational causal inference. The technologies available for that problem can in some cases be adapted for use in matching pre-randomization (Greevy et al. (2004); Ho et al. (2007)). A large number of these methods, including optimal matching, are collected in MatchIt software (Ho et al. (2009)).

## 3. MODEL VS. DESIGN-BASED ESTIMATORS FOR MATCHED PAIR EXPERIMENTS

Hill and Scott's (2009) informative commentary raises the venerable contrast between model-free and model-based estimators, to which we offer four points. First, we agree that models are sometimes warranted, valuable, or unavoidable. For example, our encompassing approach (Section 4.5 in our article) is a hierarchical model that adds modeling assumptions in order to potentially gain greater efficiency, although at the risk of greater bias; in our application, we multiply impute missing data with a model (Honaker and King (2009)); and the article on the design of our experiment proposed modeling to correct for certain types of possible experimental failures (which, as it turned out, did not materialize) (King et al. (2007)).

Second, models are sometimes useful in providing helpful intuition. For example, Hill and Scott (2009) write "In some ways, the IKN framework is actually quite similar to the multilevel framework that allows for variation in treatment effects across pairs." In fact, we prove in Section 3.2 that Hill and Scott's model without covariates is *identical* to our design-based estimator when the within-pair cluster sizes are the same. The two approaches only diverge in meaningful ways when covariates are included.

Third, randomization along with a design-based (i.e., model-free) estimator has benefits no model can match: instead of inferences that are somewhat robust to some types of model misspecification in some circumstances, design-based estimators are entirely invariant to any modeling or ignorability assumptions. This is the unique and extraordinary contribution of the idea of randomization to causal inference, when used with appropriate methods. In contrast, in even pristine experimental data, using the wrong model can generate bias, inefficiency, higher mean square error, and incorrect confidence interval coverage, especially in small samples (Freedman (2008)). While modeling can improve efficiency under some circumstances like the simulations of Hill and Scott, jettisoning the advantage of randomization by introducing unnecessary modeling assumptions is not something that should be done routinely. Although researchers who have put in the extra effort and expense, and often special Institutional Review Board approval to implement a randomized study may in some situations agree to sacrifice the guarantee of unbiasedness for a chance at lower variances, such a choice comes at substantial risk. It is no wonder that the vast majority of experimentalists, recognizing randomization as the greatest strength of their research design, abhor unnecessary assumptions and avoid model-based estimation in most stuations. See Imai et al. (2008); Imbens (2009).

Fourth, unnecessary modeling can introduce more severe biases when applied in the context of experimental failures common in real world applications. An important example of this issue occurs when controlling for a covariate that is influenced by the treatment variable, which can result in post-treatment bias. For example, in community-based experimental settings, covariates measured in baseline surveys just before the introduction of treatment may capture behavioral changes arising from subjects' anticipation of being in the control group and from other experimenter and observer intervention that may differ between treatment and control clusters, a common situation in observational studies (Ashenfelter (1978)). Indeed, the data generation process for the Monte Carlo simulations in Hill and Scott (2009) injects this real world post-treatment variable problem into the data (see Section 3.1). We show that in this situation model-based estimates are not robust to small changes in the simulation setup.

Finally, the most important risk in resorting to unnecessary modeling assumptions is the introduction of *model dependence* (King and Zeng (2006); Ho et al. (2007)). Indeed, we show analytically in Section 3.3 and via simulation in Section 3.4 that model-based inferences in experimental data can be highly model dependent. We then offer two simulated examples. In one, changing a linear modeling assumption to a nonlinear modeling assumption produces large biases and incorrect confidence interval coverage, and in such a way that model fit tests do not avoid. And in the other, we show that adjusting for a pre-treatment but incorrect covariate can produce inefficient estimates and lead to confidence intervals with inaccurate coverage when compared to the design-based estimator.

### 3.1 The Data Generation Process

We begin with Hill and Scott's (2009) data generating process. For individual $i$, in cluster $j$ ($j = 1, 2$), and pair $k$ ($k = 1, \ldots, K$), we generate individual level potential outcomes as $Y_{ijk}(t) \stackrel{\text{i.i.d.}}{\sim} \mathcal{N}(Y_{.jk}(t), \sigma_\epsilon^2)$, where $t = 1$ is treated, $t = 0$ is control. Under their data generating process,

(1) $\quad Y_{.1k}(0) \stackrel{\text{i.i.d.}}{\sim} \mathcal{N}(\mu_0, \sigma_0^2),$

(2) $\quad Y_{.2k}(0) \stackrel{\text{i.i.d.}}{=} Y_{.1k}(0) + \delta_k, \quad \delta_k \sim \mathcal{N}(0, \sigma_\delta^2),$

(3) $\quad Y_{.1k}(1) = Y_{.1k}(0) + \tau_{1k},$

(4) $\quad Y_{.2k}(1) = Y_{.2k}(0) + \tau_{2k},$

where $\mu_0$ is the mean cluster-level potential outcome under control, and $\sigma_\delta$ represents the standard deviation of within-pair imbalance. Furthermore, Hill and Scott set the causal effect (the difference in the potential outcomes, averaged over all individuals within a cluster) as $\tau_{jk} = 30/Y_{.jk}(0)$. This specification implies that $\tau_{jk}$ does not have finite moments and thus the population average treatment effect does not exist.

Hill and Scott further assume that the cluster is treated ($t = 1$) if $j = 2$ and not ($t = 0$) if $j = 1$. This means that the distributions of potential outcomes are different between the treatment and control groups, which indicates that this is a simulation where the randomization failed: Although the means of the potential outcome are the same, their variances are different unless $\sigma_\delta = 0$.

Hill and Scott then generate their cluster-level covariate as

(5) $\quad \begin{aligned} X_{jk} &= X_{jk}(T_{jk}) = Y_{.jk}(0) + \zeta_{jk} \\ &= Y_{.1k}(0) + T_{jk}\delta_k + \zeta_{jk}, \end{aligned}$

where $T_{jk}$ is the cluster-level treatment indicator and $\zeta_{jk} \stackrel{\text{i.i.d.}}{\sim} \mathcal{N}(0, \sigma_\zeta^2)$. The specification implies that $X_{jk}$ *is a post-treatment covariate* since the distribution of $X_{jk}$ is a consequence of treatment and, in particular, different between the treatment and control groups. Again, although the mean of $X_{jk}$ is the same (and equal to $\mu_0$), its variance is different unless $\sigma_\delta = 0$. Note especially that all random deviations from the normal draw of $Y_{.1k}(0)$ are reflected in $X_{jk}$, which accounts for its fit to the data.

In their simulations, the results from which we replicate exactly, Hill and Scott sample cluster sizes from a multinomial distribution with a mean of 50. In the simulations we present here, we similarly sample cluster sizes from a multinomial distribution, but vary the average cluster size to represent other typical cluster-randomized experimental settings that commonly employ fewer clusters than were used in the Seguro Popular evaluation.

### 3.2 The Model without Covariates: Equivalent to the Design-Based Estimator

Hill and Scott propose the following model without covariates, which we show here is equivalent to our design-based estimator when the within-pair cluster sizes are the same:

$$(6) \quad Y_{ijk} = \tau_k T_{jk} + \alpha_k + \epsilon_{ijk},$$

$$\text{where } \epsilon_{ijk} \stackrel{\text{i.i.d.}}{\sim} \mathcal{N}(0, \sigma_\epsilon^2),$$

$$(7) \quad \begin{pmatrix} \tau_k \\ \alpha_k \end{pmatrix} \stackrel{\text{i.i.d.}}{\sim} \mathcal{N}\left[\begin{pmatrix} \tau_0 \\ \alpha_0 \end{pmatrix}, \begin{pmatrix} \sigma_\tau^2 & \sigma_{\alpha\tau} \\ \sigma_{\alpha\tau} & \sigma_\alpha^2 \end{pmatrix}\right],$$

where $\tau_k$ is the pair-specific average treatment effect and $\epsilon_{ijk} \perp\!\!\!\perp (\tau_k, \alpha_k)$. We rewrite this model as

$$(8) \quad Y_{ijk} \mid T_{jk} \stackrel{\text{i.i.d.}}{\sim} \mathcal{N}[\tau_0 T_{jk} + \alpha_0,$$
$$\sigma_\epsilon^2 + T_{jk}(\sigma_\tau^2 + 2\sigma_{\alpha\tau}) + \sigma_\alpha^2].$$

Then, the maximum likelihood estimate of $\tau_0$ is

$$(9) \quad \hat{\tau}_0 = \frac{\sum_{k=1}^K \sum_{j=1}^2 \sum_{i=1}^{n_{jk}} T_{jk} Y_{ijk}}{\sum_{k=1}^K \sum_{j=1}^2 T_{jk} n_{jk}}$$
$$- \frac{\sum_{k=1}^K \sum_{j=1}^2 \sum_{i=1}^{n_{jk}} (1-T_{jk}) Y_{ijk}}{\sum_{k=1}^K \sum_{j=1}^2 (1-T_{jk}) n_{jk}},$$

which is identical to our design-based estimator when the within-pair cluster sizes are the same. In simulations, we find that this estimator is quite similar to the design-based estimator even when the within-pair cluster sizes are different.

### 3.3 The Model with Covariates

Consider a generalized version of the model in Section 3.2 with a covariate:

$$(10) \quad Y_{ijk} = \alpha_k + g(X_{jk})\beta + \tau_k T_k + \epsilon_{ijk},$$

$$\text{where } \epsilon_{ijk} \stackrel{\text{i.i.d.}}{\sim} \mathcal{N}(0, \sigma_\epsilon^2),$$

where $g(\cdot)$ is an assumed function specified as part of the model and $(\tau_k, \alpha_k)$ is distributed as the bivariate normal in equation (7). Hill and Scott (2009) consider a special case of this model with a post-treatment covariate [see equation (5)], such that

$$(11) \quad \begin{aligned} g(X_{jk}) &= g(X_{jk}(T_{jk})) = g(Y_{.jk}(0)) \\ &= g(Y_{.1k}(0) + T_{jk}\delta_k + \zeta_{jk}), \end{aligned}$$

and with the linear functional form restriction, $g(x) = x$.

If we estimate this general model using Hill and Scott's post-treatment covariate, the crucial question is what quantity is being estimated. We denote this estimand as $\tau^*$ and characterize the difference between it and the average treatment effect (under this model) as follows (see Rosenbaum (1984)):

$$\tau^* - E(\tau_k)$$
$$(12) \quad \equiv E\{E(Y_{ijk} \mid T_{jk} = 1, X_{jk})$$
$$- E(Y_{ijk} \mid T_{jk} = 0, X_{jk})\} - E(\tau_k),$$
$$(13) \quad = E\{E(Y_{ijk} \mid T_{jk} = 1, X_{jk}(1))$$
$$- E(Y_{ijk} \mid T_{jk} = 0, X_{jk}(0))\} - E(\tau_k),$$
$$(14) \quad = E\{g(Y_{.1k}(0) + \delta_k + \zeta_{jk})$$
$$- g(Y_{.1k}(0) + \zeta_{jk})\}\beta.$$

The model dependence of Hill and Scott's specification can be seen in the last line: When $g(x) = x$ as they assume, then the last line equals 0 and discrepancy between the estimand and the quantity of interest vanishes. However, if $g(\cdot)$ is not a linear function, then the quantity being estimated by this model, $\tau^*$, does not in general equal the average treatment effect, that is, $E(\tau_k)$. The degree of discrepancy thus solely depends on the functional form assumption, which of course is a clear case of model dependence.

### 3.4 Simulations

We perform two simulations which are based on, but not identical to, Hill and Scott's simulation setup. Our goal in this section is to offer a more general illustration of model dependence than in Hill and Scott's setup. To do so, in both simulations, we correct the randomization failure by properly randomizing the treatment and address the dvivide-by-zero problem by using a left-truncated normal distribution (instead of a normal distribution without truncation) with a truncation point of 2. We then examine the consequence of adjusting for the post-treatment variable (the first simulation) as well as the pre-treatment variable (the second simulation). We run our simulation for 2,000 iterations; details appear in our replication data archive (Imai et al. (2009a)).

For the first simulation, we set $g(X_{jk}) = \log(Y_{.1k}(0)) + \eta_{jk}$ if $T_{jk} = 0$ and $g(X_{jk}) = \exp(Y_{.2k}(0)) + \eta_{jk}$ if $T_{jk} = 1$, where $\eta_{jk} \overset{\text{i.i.d.}}{\sim} \mathcal{N}(0,2)$, the values of which are fixed over simulations. The results of this first simulation appear in the left column of Figure 1. The horizontal axis for each graph is the standard deviation of added (post-treatment) imbalance, which is denoted by $\sigma_\delta$ (see Section 3.1). We present results for our design-based estimator (solid line), the model-based estimator with a covariate (dashed line), and—to evaluate whether it might be possible to test one's way out of the problems—a pre-test model-based estimator using the likelihood ratio test to decide for each simulation whether to include the covariate, as done in Hill and Scott's simulations (dotted line).

The estimated bias presented in the top left graph shows that while our design-based estimator is approximately unbiased, the model-based and pre-test estimators are severely biased for a wide range of the simulations. The variance of each of the estimators is relatively small, and so with large bias the root mean square error is mostly irrelevant, but it too indicates (in the middle left graph) that the design-based estimator is superior. The estimated coverage probability of the 95% confidence interval, displayed for the two estimators in the bottom left graph, stays approximately at the nominal level for our design-based estimator but is far from valid over much of the range for the model-based and pre-test approaches.

For our second set of simulations, we examine the consequence of adjusting for the *pre-treatment* covariate using a model-based approach. We adopt a data generating process similar to that of Hill and Scott's simulations, but use a different specification for the pre-treatment cluster-level covariate; $X_{1k} = \log(Y_{.1k}(0)) + \eta_{jk}$ and $X_{2k} = \exp(Y_{.2k}(0)) + \eta_{jk}$, where $\eta_{jk} \overset{\text{i.i.d.}}{\sim} \mathcal{N}(0,2)$. In addition, because many community-based cluster-randomized experiments in public health and education are forced to use as few as 5 to 10 pairs, we reduced the sample size to twenty clusters of average size 15.

The results from this second simulation appears in the right column of Figure 1, again for design-based (solid line), model-based (dashed) and pre-test (dotted) estimtors. As expected, the top right graph shows that all three estimators are approximately unbiased because we no longer adjust for post-treatment covariates (although the bias is slightly smaller for the pre-test and design-based estimators than the model-based approach). The middle right graph shows that the design based estimator has uniformly lower root mean square error than the other two approaches. The bottom right graph shows that our design-based approach produces approximately correct coverage across varying levels of within-pair imbalance, while the model-based and pre-test estimators produce confidence intervals that are somewhat too narrow.

### 3.5 How to Use Pre-Treatment Information

Introducing models into randomized experiments can improve estimation or make it worse. Hill and Scott have given examples where specific models out-perform design-based estimators. With similar models and data generation processes we show here that models can also under-perform relative to design-based estimators. Although diagnostic tests can sometimes help an analyst choose the correct strategy from the data, the differences can be subtle and in many situations, such as the ones we illustrate here, standard tests cannot detect model failures. None of these points are new, but it is useful to have examples of each issue laid out with the clarity this Symposium has made possible.

Given these issues, our recommendation, along with, it seems, our discussants, is to avoid modeling choices by using pair matching as part of the design of cluster-randomized experiments on all available covariates *prior to* randomization. This allows researchers to obtain efficiency gains of modeling without risking the statistical advantages of random assignment. If exact pair matching is possible, then model dependence is eliminated and the difference between many model-based and design-based estimators will vanish. When exact matching is not possible, then the user may choose to introduce a model if the risks of that approach are not outweighed by the benefits of guaranteed unbiasedness due to randomization. In many cases, such as with noncompliance and missing data, models may be unavoidable.

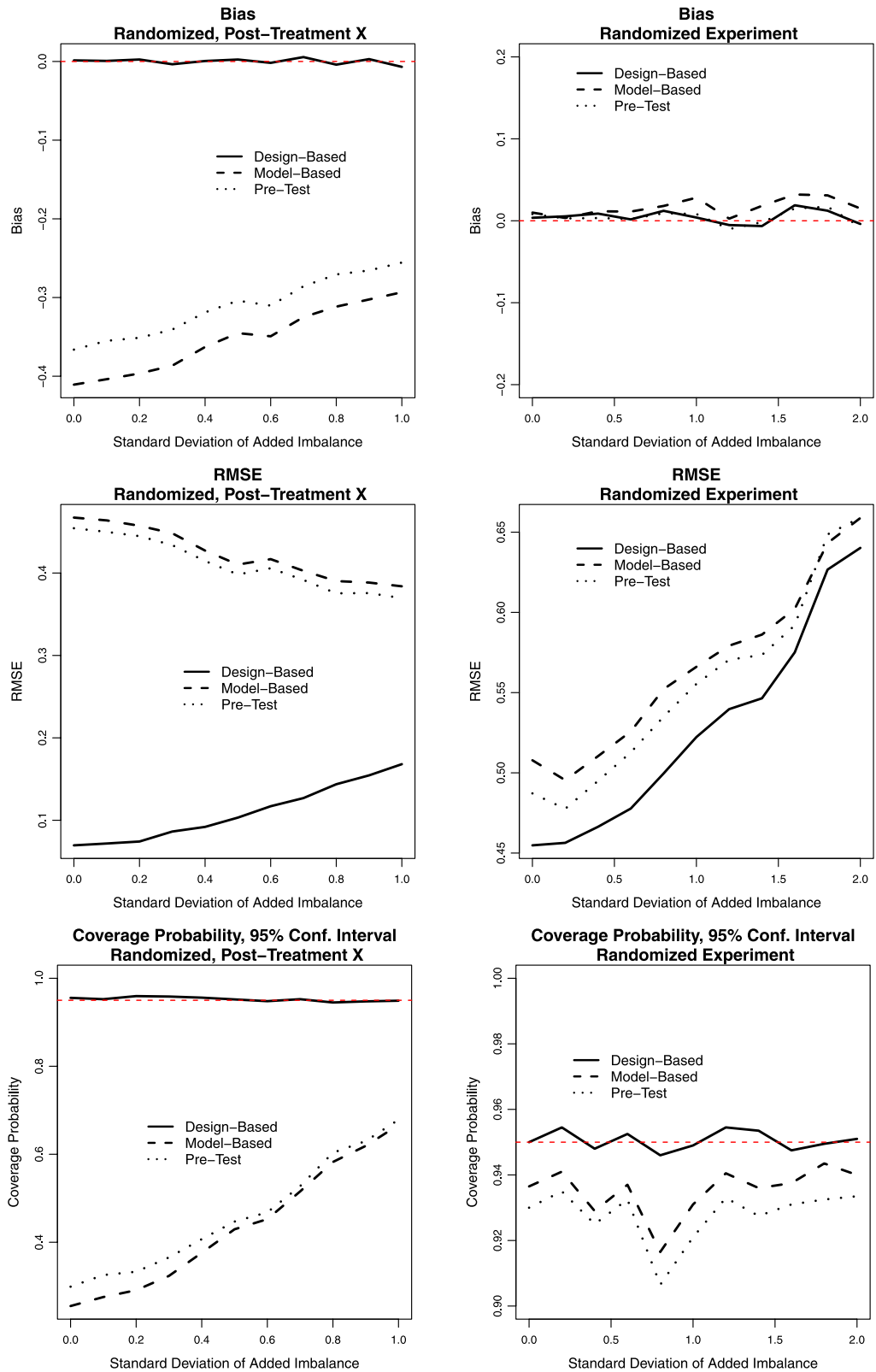

FIG. 1. *Model dependence. For the design-based (solid), model-based (dashed) and pre-test (dotted) estimators, we present the bias (top row), root mean square error (middle row) and confidence interval coverage (bottom row). The left column demonstrates model dependence from the simulation in Hill and Scott by changing only the model to add nonlinearity; the right column gives an example where even under proper randomization inclusion of a covariate can worsen RMSE and the coverage probability.*

## 4. CONCLUDING REMARKS

We developed the arguments, methods and evidence for our article in the context of a large randomized study of the Mexican universal health care system, *Seguro Popular* (King et al. (2007), 2009). Using the matched pair design for cluster randomization and our design-based statistical methods means that we were able to save a great deal of money and produce far more informative causal effects without risky assumptions. As our discussants have made clear, these results should be widely applicable, and the matched pair design should be used whenever feasible. Fortunately, in cluster-randomized studies, matching clusters in pairs usually is feasible, at least much more so than for some classes of unit-randomized studies. As our content analysis of the scholarly literature shows, there is much room for improvement in the practice of experimental design; this symposium offers a clear path to saving research resources and unearthing far more information, in cluster randomized experiments, than has been understood heretofore.

We thank our discussants again for their informative contributions, and we look forward to many applications across fields of inquiry, as well as new research that pushes forward experimental design in ways that continue to make possible more scientifically valid and efficient public policy evaluations.

## APPENDIX: JOURNALS INCLUDED IN CONTENT ANALYSIS

We included journals in the content analysis reported in Table 1 if they published at least one cluster-randomized trial during the study period, which was 2003–2009 for political science and 2006–2009 for the others. The journals included are as follows.

Medicine and public health: *American Journal of Public Health*, *American Journal of Sports Medicine*, *Annals of Internal Medicine*, *British Medical Journal*, *Journal of the American Medical Association*, *Lancet*, *Medicine & Science in Sports & Exercise*, *New England Journal of Medicine*. Economics: *American Economic Review*, *Econometrica*, *Journal of Political Economy*, *Journal of Policy Analysis and Management*. Education: *American Education Research Journal*, *American Journal of College Health*, *Educational Evaluation and Policy Analysis*. Political science: *American Behavioral Scientist*, *American Journal of Political Science*, *American Political Science Review*, *American Politics Research*, *Annals of the American Academy of Political and Social Science*, *Comparative Political Studies*, *Electoral Studies*, *Journal of Politics*, *Political Analysis*, *Political Psychology*, *Political Research Quarterly*, and *PS: Political Science and Politics*.

## ACKNOWLEDGMENTS

Many thanks to our discussants for their time, attention and contributions, and to Jennifer Hill and Marc Scott for providing their data and replication information. Financial support is provided in part by NSF Grant SES-07-52050.

## REFERENCES

Ashenfelter, O. (1978). Estimating the effect of training programs on earnings. *Rev. Econ. Statist.* 47–57.

Freedman, D. A. (2008). On regression adjustments to experimental data. *Adv. in Appl. Math.* **40** 180–193. MR2388610

Greevy, R., Lu, B., Silver, J. H. and Rosenbaum, P. (2004). Optimal multivariate matching before randomization. *Biostatistics* **5** 263–275.

Hill, J. and Scott, M. (2009). Comment on "The essential role of pair matching." *Statist. Sci.* **24** 54–58.

Ho, D., Imai, K., King, G. and Stuart, E. (2007). Matching as nonparametric preprocessing for reducing model dependence in parametric causal inference. *Political Analysis* **15** 199–236. Available at http://gking.harvard.edu/files/abs/matchp-abs.shtml.

Ho, D. E., Imai, K., King, G. and Stuart, E. A. (2009). Matchit: Nonparametric preprocessing for parametric causal inference. *J. Statist. Software*. To appear. Available at http://gking.harvard.edu/matchit.

Honaker, J. and King, G. (2009). What to do about missing values in time series cross-section data. Available at http://gking.harvard.edu/files/abs/pr-abs.shtml.

Iacus, S. M., King, G. and Porro, G. (2008). Matching for causal inference without balance checking. Available at http://gking.harvard.edu/files/abs/cem-abs.shtml.

Imai, K., King, G. and Nall, C. (2009a). Replication data for: The essential role of pair-matching in cluster-randomized experiments, with application to the mexican universal health insurance evaluation: Rejoinder. Available at hdl:1902.1/12730 UNF:3:CKs4T0iVYxP36LQSMgAkuw== Murray Research Archive [Distributor].

Imai, K., King, G. and Nall, C. (2009b). The essential role of pair matching in cluster-randomized experiments, with application to the mexican universal health insurance evaluation. *Statist. Sci.* **24** 29–53.

Imai, K., King, G. and Stuart, E. (2008). Misunderstandings among experimentalists and observationalists about causal inference. *J. Roy. Statist. Soc. Ser. A* **171** 481–502. Available at http://gking.harvard.edu/files/abs/matchse-abs.shtml. MR2427345

Imbens, G. (2009). Better LATE than nothing: Some comments on Deaton (2009) and Heckman and Urzua (2009). Working paper, NBER.

King, G., Gakidou, E., Imai, K., Lakin, J., Moore, R. T., Nall, C., Ravishankar, N., Vargas, M., Téllez-Rojo, M. M., Ávila, J. E. H., Ávila, M. H. and Llamas, H. H. (2009). Public policy for the poor? A randomised assessment of the Mexican universal health insurance programme. *The Lancet*. To appear. Available at http://gking.harvard.edu/files/abs/spi-abs.shtml.

King, G., Gakidou, E., Ravishankar, N., Moore, R. T., Lakin, J., Vargas, M., Téllez-Rojo, M. M., Ávila, J. E. H., Ávila, M. H. and Llamas, H. H. (2007). A 'politically robust' experimental design for public policy evaluation, with application to the Mexican universal health insurance program. *J. Policy Anal. Manag.* **26** 479–506. Available at http://gking.harvard.edu/files/abs/spd-abs.shtml.

King, G. and Zeng, L. (2006). The dangers of extreme counterfactuals. *Political Anal.* **14** 131–159. Available at http://gking.harvard.edu/files/abs/counterft-abs.shtml.

Rosenbaum, P. (1984). The consequences of adjusting for a concomitant variable that has been affected by the treatment. *J. Roy. Statist. Soc. Ser. A* **147** 656–666.

Zhang, K. and Small, D. S. (2009). Comment on "The essential role of pair matching in cluster-randomized experiments, with application to the mexican universal health insurance program." *Statist. Sci.* **24** 59–64.